\documentclass[aps,pra,reprint,showpacs,twocolumn,longbibliography,superscriptaddress,nofootinbib]{revtex4-1}
\usepackage{graphicx}
\usepackage{dcolumn}
\usepackage{bm}
\usepackage{braket}
\usepackage{leftidx}
\usepackage{cancel}
\usepackage{amsmath}
\usepackage{epstopdf}
\usepackage{color}
\usepackage[mathcal]{eucal}
\usepackage{float}

\begin{document}
\title
{\bf  Single-molecule-resolution ultrafast near-field optical microscopy \\ via plasmon lifetime extension}

\author{Rasim Volga Ovali}
\affiliation{Department of Physics, Recep Tayyip Erdogan University, 53100 Rize, Turkey}

\author{Ramazan Sahin}
\affiliation{Faculty of Science, Department of Physics, Akdeniz University, 07058 Antalya, Turkey}

\author{Alpan Bek}
\affiliation{Department of Physics, Middle East Technical University, 06800 Ankara, Turkey}
\author{Mehmet Emre Tasgin} \thanks{metasgin@hacettepe.edu.tr }
\affiliation{Institute  of  Nuclear  Sciences, Hacettepe University, 06800 Ankara, Turkey}

\date{\today}

\begin{abstract}
A recent study shows that: when a long lifetime particle is positioned near a plasmonic metal nanoparticle, lifetime of plasmon oscillations extends, but, \textit{only} near that long-life particle~[PRB 101, 035416 (2020)]. Here, we show that this phenomenon can be utilized for ultrahigh~(single-molecule) resolution ultrafast apertureless~(scattering) SNOM applications. We use the exact solutions of 3D Maxwell equations. We illuminate a metal-coated silicon tip, a quantum emitter~(QE) placed on the tip apex, with a femtosecond laser. The induced near-field in the apex decays rapidly except in the vicinity of the sub-nm-sized QE. Thus, the resolution becomes solely limited by the size of the QE. As positioning of a QE on the tip apex is challenging, we propose the use of a newly-discovered phenomenon; stress-induced defect formation in 2D materials. When a monolayer, e.g., transition metal dichalcogenide~(TMD) is transferred to the AFM tip, the tip indentation of 2D TMD originates a defect-center located right at the sharpest point of the tip; that is exactly at its apex. Moreover, the resonance of the defect is tunable via a voltage applied to the tip. Our method can equally be used for background-noise-free nonlinear imaging and for facilitating single-molecule-size chemical manipulation.
\end{abstract}
\maketitle

Metal nanoparticles~(MNPs) interact very strongly with incident optical light. Surface plasmons, oscillations of free electrons, localize the incident light into nm-sized hotspots. Hotspot intensity can be 5 to 7 orders of magnitude larger than  one of the incident light~\cite{hoppener2012self}. Hotspots can enhance both linear~\cite{LavrinenkoPRB2019,Cherenkov_Nanophotonics_2020} and nonlinear optical  response~\cite{kauranen2012nonlinear}. Enhanced hotspot intensity allows detection of even a single molecule~\cite{betzig1993single}.  In a nonlinear process both incident and converted fields are enhanced. Thus, a nonlinear process (for instance Raman scattering) can be enhanced quadratically~\cite{kauranen2012nonlinear}.

Localization of light not only strengthens the near-fields, but also enables a hotspot-size resolution at optical imaging~\cite{bazylewski2017review} and surface manipulation~\cite{zhan2018plasmon}. Hotspot of a metal-coated atomic force microscope~(AFM) tip can provide surface images with resolution down to $\sim$10 nm ~\cite{alpan1,bazylewski2017review}, limited by the tip apex size, at optical and near-infrared wavelengths. This method, scanning near-field optical microscopy~(SNOM), is essential in particular for investigations of light-matter interaction in strongly-correlated condensed phase and 2D materials~\cite{liu2016nanoscale}. This is because, amplitude/phase of the light scattered by the probe (tip) contains information about the local dielectric function of the scanned material~\cite{hillenbrand2000complex}. For instance; structures of different phases in VO$_2$~\cite{qazilbash2007mott,liu2013anisotropic} and organic thin films (for energy harvesting)~\cite{westermeier2014sub}, plasmon mode profiles~\cite{chen2012optical,fei2012gate,hauer2015solvothermally} and exciton-polariton transport in 2D materials~\cite{hu2017imaging,fei2016nano} can be imaged via apertureless SNOM~(a-SNOM) (also referred  as scattering-type SNOM or s-SNOM)~\cite{chen2019modern}. SNOM has numerous applications in physics, biology, chemistry and engineering~\cite{amenabar2013structural,dunn1999near}.

Although former experiments with SNOM confined to continuous-wave~(cw) sources, recent studies are focused on femtosecond~(fs) optical/infrared sources~\cite{yao2020nanoimaging} which resolve the ultrafast transient spatiotemporal dynamics, for instance, of exciton-polariton dynamics in 2D materials~($\sim$10 fs, $\sim$10 nm)~\cite{mrejen2019transient,steinleitner2017direct,ni2016ultrafast,wagner2014ultrafast,kravtsov2016plasmonic} and solid state phase transition~\cite{donges2016ultrafast} far from equilibrium. Ultrafast SNOM enabled the observation of intriguing physics, for instance, negative ``phase velocity" for phonon-polaritons in 2D hexagonal boron nitride~\cite{yoxall2015direct}. Such a fs time-resolved imaging of electric fields is important for tracing ultrashort processes taking place in biology and chemistry~\cite{martin2004femtochemistry}. Limitations on pulse repetition rate and tip oscillation frequencies, faced in initial experiments, are circumvented by correlating the tip oscillations with the tip-scattered field~\cite{wang2016scattering}.

Spatial resolution of SNOM, however, is still limited with the tip apex size~($\sim$10 nm)~\cite{mrejen2019transient,steinleitner2017direct,ni2016ultrafast,wagner2014ultrafast,kravtsov2016plasmonic}. Furthermore, manufacturing of metal-coated AFM tip of apex size $\sim$10 nm is quite challenging. Fortunately, here we show  that findings of a recent study~\cite{yildiz2020plasmon} can overcome the challenges faced in the spatial resolution, intriguingly, via utilizing a phenomenon in its temporal dynamics. The method, described below, relies on the enhanced lifetime of charge oscillations taking place ``only" near a quantum object~(QO)~\cite{PS_QO}.

%
%
%

Dark-hot resonances~\cite{stockman2010dark}, also referred as Fano resonances~\cite{luk2010fano,limonov2017fano}, appear when a bright-mode is coupled to a longer-life dark plasmon mode~\cite{tassin2012electromagnetically}. The local field can be further enhanced via extended lifetime of bright plasmons~\cite{yildiz2020plasmon}, on top of the enhancement due to localization. A nonlinear process, e.g., Raman scattering, can be further enhanced (again quadratically) by aligning both incident and converted fields with two dark-hot (Fano) resonances~\cite{he2016near,ye2012plasmonic}. A similar lifetime enhancement enable the operation of spasers ---metal nanoparticles~(MNPs) coated with molecules~\cite{noginov2009demonstration}. A Fano resonance, demonstrating a dip in its steady-state excitation spectrum~\cite{limonov2017fano,TasginFanoBook2018}, ironically, enhances the plasmon energy accumulation in its temporal dynamics~\cite{stockman2010dark,yildiz2020plasmon}.

Ref.~\cite{yildiz2020plasmon} demonstrates that surface charge oscillations of a MNP near a longlife nanomaterial last much longer than the lifetime of the bright plasmon. As the lifetime extends only near the long lifetime nanomaterial, it shows itself as a weak narrowing in the absorption spectrum~\cite{yildiz2020plasmon}. This is unlike a spaser whose surface is covered completely with a vast number of molecules~\cite{noginov2009demonstration}. 

\begin{figure}
\centering
\includegraphics[width=0.47\textwidth,trim={0 0 0 0},clip]{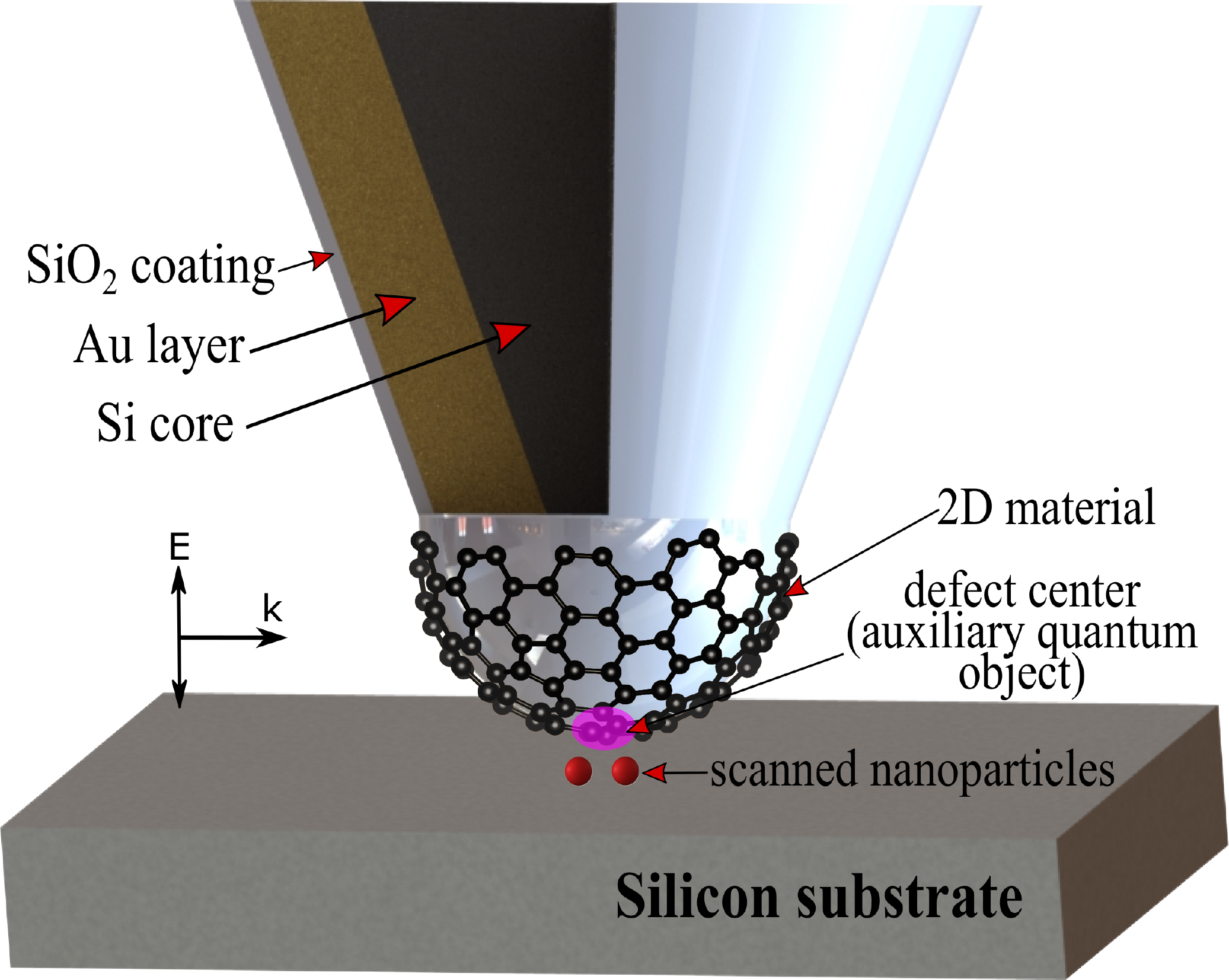} 
\caption{A 10nm-thick silicon AFM tip is coated with 10nm gold. Gold surface is also coated with a 1.5nm insulator (${\rm SiO}_2$) layer to avoid charge transfer between the quantum object and gold surface. A 2D material is grown on the tip, with maximum bending at the tip apex, creates a stress-induced excitonic defect-center which scans two 2-nm-sized nanoparticles. Tip is excited by a z-polarized 5 fs ultrashort pulse propagating in the x-direction. While the near-field excitations decay in 13.7 fs on the gold surface, near-field on the QO lasts much longer, i.e. 240 fs, see Fig. 3. This allows the utilization of the setup as a single-QO-sized resolution SNOM and chemical manipulation. }
\label{fig1}
\end{figure}

\begin{figure}
\centering
\includegraphics[width=0.47\textwidth,trim={0 0 0 0},clip]{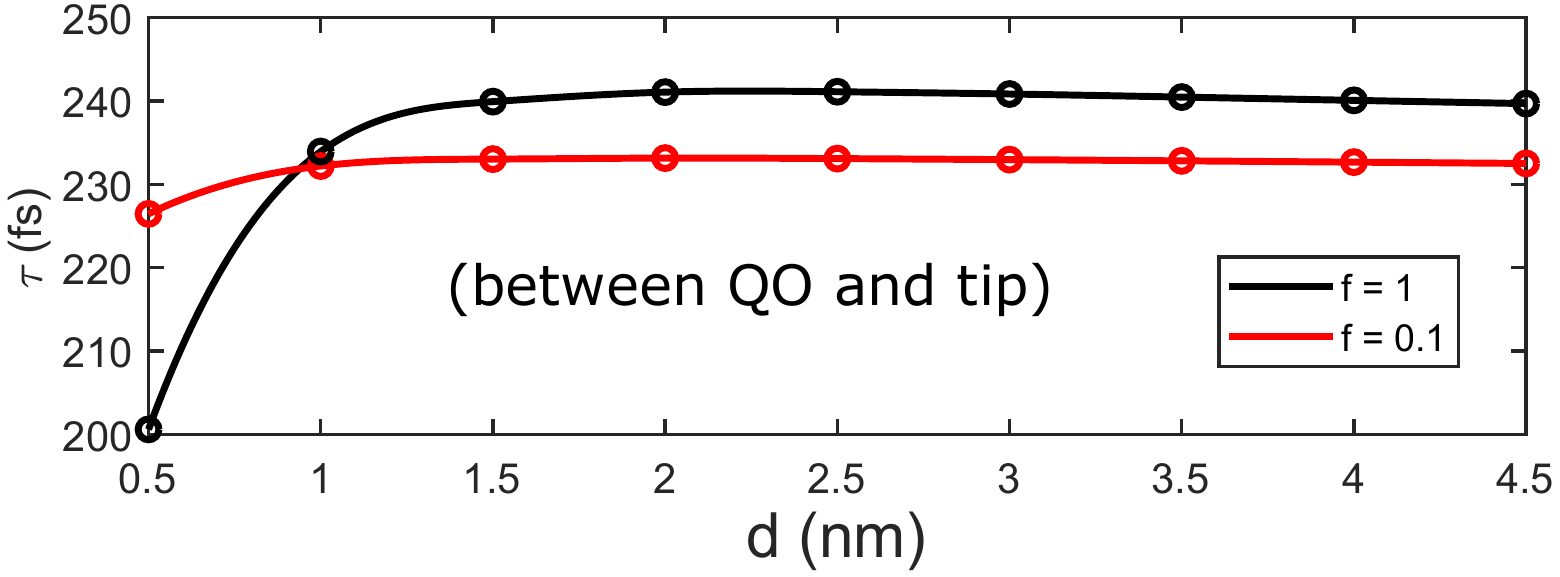} 
\caption{Lifetime of electric field oscillations ``between" the tip apex and the auxiliary QO when the system is illuminated with a 5 fs pulse. Oscillator strength of the auxiliary QO is chosen as $f=1$, corresponding to a stress-induced excitonic defect center in a 2D material, and $f=0.1$ for a molecule~\cite{wu2010quantum,premaratne2017theory}. }
\label{fig2}
\end{figure}

\begin{figure}
\centering
\includegraphics[width=0.47\textwidth,trim={0 0 0 0},clip]{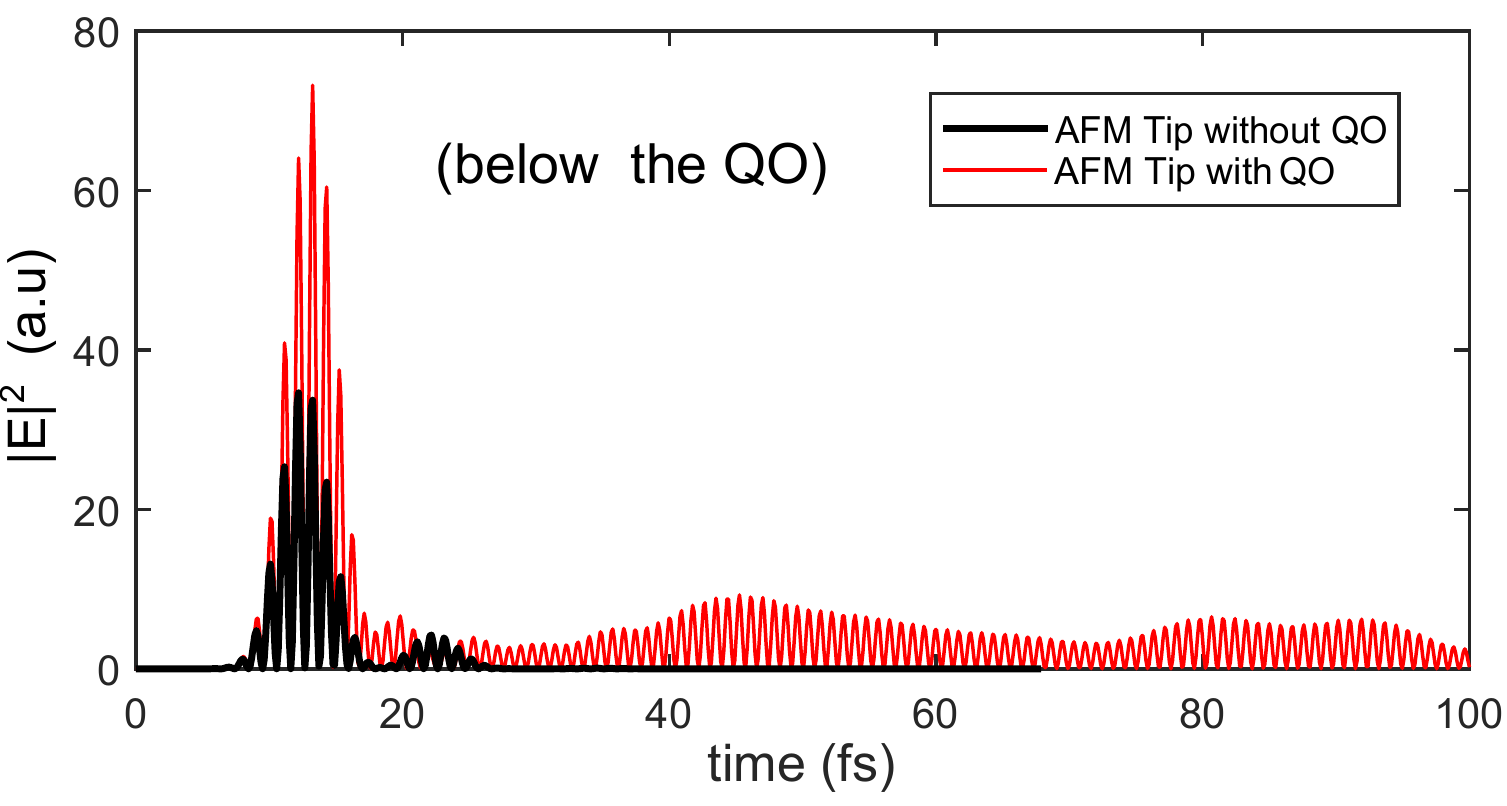} 
\caption{(black) Time evolution of the electric field on the gold surface, at the tip apex, when there is no QO. Oscillations decay quickly after $\sim$25 fs. (red) An auxiliary QO is utilized to extend the lifetime of plasmonic oscillations. Electric field ``0.25 nm-below" the auxiliary QO lasts much longer.}
\label{fig3}
\end{figure}

\begin{figure}
\centering
\includegraphics[width=0.47\textwidth,trim={0 0 0 0},clip]{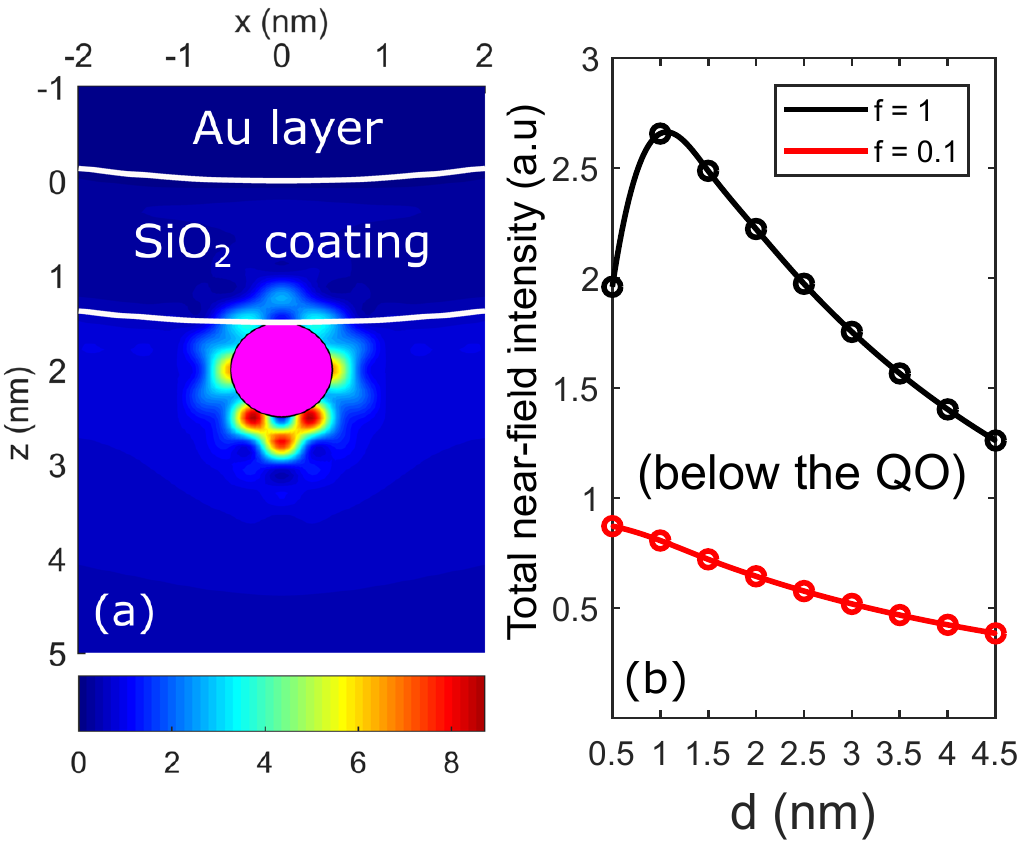} 
\caption{ (a) Total near-field intensity, $J_{\rm near}({\bf r})=\int dt \: I_{\rm near}({\bf r},t)$, scanning the two 2-nm sized nanoparticles in Fig.~\ref{fig1}, is confined around the QO. The fringe-like behavior appears because the 5 fs pulse contains a broadband frequency. Hence, both dipole and quadrupole modes are excited~\cite{PS2}. (c) Total near-field intensity, just below the QO, for different distances between the QO and the gold surface.}
\label{fig4}
\end{figure}

\begin{figure}
\centering
\includegraphics[width=0.47\textwidth,trim={0 0 0 0},clip]{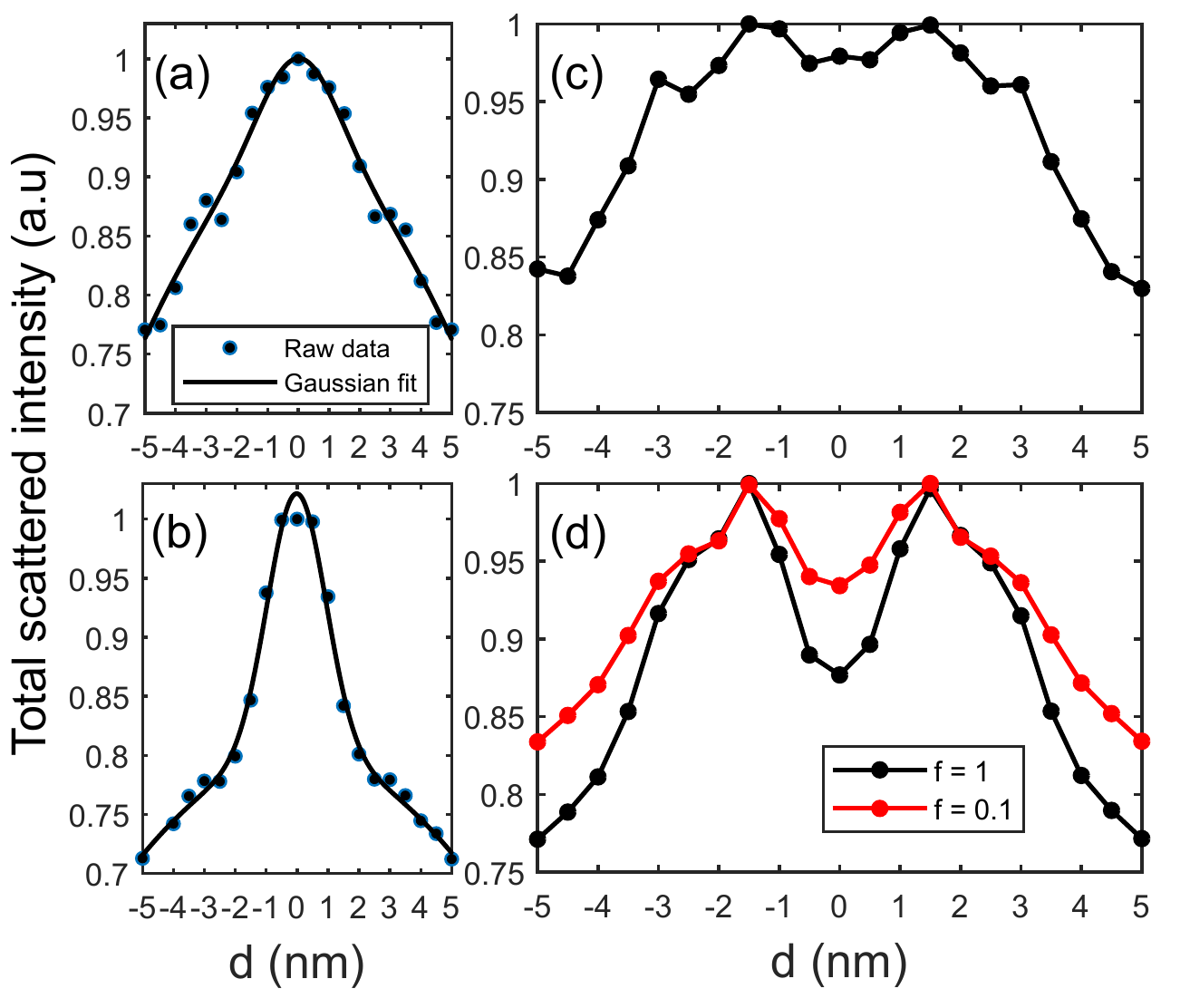} 
\caption{AFM tip scans (a,b) a single 2 nm-sized nanoparticle and (c,d) two 2 nm-sized nanoparticles separated 2 nm from each other.  Total scattered intensity in the (upper) absence and (lower) presence of the auxiliary QO. The tip apex is 30 nm. The \textit{deceptive} high-resolution of nanoparticles in (a,c) is due to the perfect hemispherical body of the tip. (d) The auxiliary QO is very successful in resolving the two nanoparticles, while the gold tip without the QO fails despite a perfectly-shaped tip.}
\label{fig5}
\end{figure}

In this paper, we utilize this phenomenon for achieving a quantum object~(QO)-sized resolution ultrafast SNOM. We illuminate a gold-coated tip with a 5 fs-long laser pulse, Fig.~\ref{fig1}. First, (i) we show that electric field of a gold-coated AFM tip lasts $\sim$17 times longer compared to a plasmon lifetime, but, {\it only near a} (auxiliary) {\it quantum object} attached to the tip apex, see Figs.~\ref{fig1} and \ref{fig2}. We perform FDTD simulations of 3D Maxwell equations and use a Lorenzian dielectric for the QO~\cite{LorentzianTMD_Eps2018,wu2010quantum,premaratne2017theory}. Field below the 1 nm-sized auxiliary QO, which scans two 2 nm-sized nanoparticles~(for instance can be proteins), also lasts that much longer, see Fig.~\ref{fig3}.  Second, (ii) we record ``total intensities"~(or fluorescence) scattered by the two 2 nm-sized nanoparticles. We show that lifetime-enhanced near-field below the auxiliary quantum object~(Figs.~\ref{fig3} and \ref{fig4}) can resolve the 2 nm-sized particle successfully~(see Fig.~\ref{fig5}). In our simulations we use a 30 nm-thick tip apex.

Positioning a single quantum emitter at the gold-coated tip apex, although possible within current nanotechnological techniques~\cite{schell2014scanning}, can be challenging. Thus, here we propose to employ a ``strain-induced" quantum defect center of a bended 2D materials, e.g., transition metal dichalcogenide~(TMD)~\cite{AtatureNatureComm2017,BendingDefectAPL2019}. When such a 2D material is  transferred to an AFM tip, the defect sits at the sharpest (bended) edge of the tip~\cite{BendingDefectAPL2019}, that is at its apex, see Fig.~1. In our approach, the QO and hence the plasmon lifetime extension is generated exactly where it is needed, at the tip apex, which is the part of the tip that actually is responsible for image formation. 

Strain-induced defects in TMDs create spatially and spectrally isolated (quantum emitter) centers for localized~(0-dimensional) excitons~\cite{Branny2017,AtatureNatureComm2017,BendingDefectAPL2019}. Excitonic nature~\cite{Moody2016,Arora2015,Gerber2019,Schneider2018,Wang2016} of the centers enables strong near-field coupling~\cite{schell2014scanning,tripathi2018spontaneous,lee2017near,bao2015visualizing} and highly-enhanced narrow bandwidth emission~\cite{Cotrufo2019,AtatureNatureComm2017,Deng2018,Feng2012,He2015,Dang2020,Branny2017,Dhakal2017} with single-photon characteristics~\cite{AtatureNatureComm2017,BendingDefectAPL2019}; which makes them perfect candidates for imaging applications~\cite{AtatureNatureComm2017,BendingDefectAPL2019}. Such QEs are already shown to demonstrate the spontaneous emission enhancement near metallic surfaces~\cite{tripathi2018spontaneous}, reverse of the lifetime enhancement~\cite{yildiz2020plasmon}, which were shown for molecules or quantum dots~\cite{anger2006enhancement} previously. Moreover, defect resonance can be controlled with an external voltage~\cite{chakraborty2015voltage,schwarz2016electrically} which can change the operation wavelength of the locally lifetime-enhanced device slightly. That is, such a setup is a perfect business for high-resolution ultrafast SNOM imaging we introduce here. The new method has the potential to trace electric field evolution of, for instance, protein bioactivities~\cite{martin2004femtochemistry} at $\sim$fs and 1 nm spatiotemporal resolution at different frequencies. For completeness, we also perform similar simulations for a $f=0.1$ QO~\cite{wu2010quantum}, a typical value for molecules/proteins~\cite{wu2010quantum,premaratne2017theory}, placed on tip apex.

The setup we study here can also be utilized for nm-size chemical manipulation~\cite{zhan2018plasmon,kazuma2020single} and in  Fano-enhanced nonlinear processes~\cite{ye2012plasmonic,postaci2018silent,zhang2013coherent}.

{\bf \small Setup.}--- We calculate the exact solutions of 3D Maxwell equations for a sample setup depicted in Fig.~\ref{fig1}. We consider a 10 nm-thick silicon AFM tip coated with 10 nm gold for plasmonic operations~\cite{mrejen2019transient,steinleitner2017direct,ni2016ultrafast,wagner2014ultrafast,kravtsov2016plasmonic}. Gold surface is also coated with a 1.5 nm insulator (Si${\rm O }_2$) layer to avoid charge-transfer between the gold surface and the auxiliary quantum object. A 2D material, displaying semiconductor feature along its body, is grown on the tip~\cite{BendingDefectAPL2019} as demonstrated in Fig.~\ref{fig1}. The bending of the 2D material creates a stress-induced defect center~\cite{AtatureNatureComm2017,BendingDefectAPL2019}. The stress-induced defect center appears at the most-curved position (i.e. apex) of the tip~\cite{BendingDefectAPL2019}. The defect, interacting strongly with the near-field, possesses a strong dipole-moment with a typical oscillator strength of $f\sim 1$. The AFM tip scans the two nanoparticles, of diameter 2 nm, sitting on a silicon substrate. We carry out 3D FDTD simulations of the setup using the experimental dielectric functions of the materials, i.e., for silicon, ${\rm SiO}_2$, and gold~\cite{johnson}. We use a Lorentzian dielectric function,  $\epsilon(\omega)=1 - f\omega_0^2/(\omega_0^2+i2\gamma\omega-\omega^2)$, for the auxiliary quantum object~\cite{LorentzianTMD_Eps2018,wu2010quantum,premaratne2017theory}. $\omega_0$ and $\gamma$ are the resonance and decay rate of the auxiliary quantum object.

Lifetime of such defect centers, in the order of $\sim$ns~\cite{BendingDefectAPL2019}, are quite long compared to the plasmonic oscillations ($\sim$10 fs in gold). So, it is possible to extend the lifetime of the near-field~(plasmon) oscillations between the defect~(QO) and the gold surface as in Ref.~\cite{yildiz2020plasmon}. Such a lifetime extension, we expect and observe here, in Fig.~\ref{fig2}, is the analog of the phenomenon observed in Ref.~\cite{yildiz2020plasmon}. Ref.~\cite{yildiz2020plasmon} employs longer lifetime of dark plasmon modes which can be $\sim$10-100 times longer compared to bright plasmon modes. Here, we employ the quantum object as the longlife particle whose lifetime is $\sim$ ns, which is 5-orders of magnitude larger than the one for plasmon oscillations on the gold surface ($\sim$10 fs).

{\bf \small Lifetime extension.}--- In Fig.~\ref{fig2}, we show that lifetime of plasmon oscillations ``between" the auxiliary QO and the gold surface~($\sim$240 fs) increases $\sim$17 times copared to the lifetime of localized surface plasmons~(LSPs) in the bare gold surface~($\simeq$13.7 fs). The plasmonic near-field oscillations out of the QO, in contrast, decay within 13.7 fs.

Lifetime extension phenomenon, employing an auxiliary quantum object, displays a common behavior with the one employing longer-life dark plason modes~\cite{yildiz2020plasmon}. Enhancement increases for closer positioning (stronger coupling) of the auxiliary QO to the gold surface. But after a critical (strong) coupling, lifetime extension decreases, a feature also observed in fluorescence enhancement of molecules near plasmonic surfaces~\cite{anger2006enhancement}.

{\bf \small Utilization.}--- Therefore, such a setup (Fig.~\ref{fig1}) is possible to be utilized as a single-QO-size resolution SNOM. For this utilization, however, a similar lifetime extension is needed to be observed also \textit{below} the auxiliary QO, the location which scans the two 2 nm-sized nanoparticles sitting on the substrate. 

In Fig.~\ref{fig3}, we plot the temporal evaluation of the electric field, just (0.25 nm) below the QO. We simulate the auxiliary QO as a 1 nm-sized sphere of Lorentzian width $\gamma=10^9$Hz, resonance $f_0=491$ THz and oscillator strength $f=1$. We illuminate the setup (Fig.~\ref{fig1}) with a z-polarized, 5 fs-width, ultrashort pulse propagating in the x-direction. We set the carrier frequency of the ultrashort pulse to $f_c=491$ THz~($\lambda_c=610$ nm), the resonance of the auxiliary QO.

We kindly remark that in this \textit{prominent} work we aim to demonstrate how the local lifetime enhancement method works for imaging implementations. Other spectral regions, e.g., 610-680~nm and 740-820~nm~\cite{AtatureNatureComm2017,BendingDefectAPL2019}, can also be used for nm-size-resolution imaging.

Fig.~\ref{fig3} shows that electric field oscillations, also below the QO, last much longer compared to the surface of a gold tip. So, near-field below the QO scan the two nanoparticles much longer times compared to another position on the gold surface (even of the tip apex would be a flat one). While the near-field of gold surface decays, after $\sim 13.7$ fs, the 1 nm-thick QO continues to provide scattering signal from the two nanoparticles. Therefore, one can achieve $\sim 1$ nm-size resolution SNOM either (i) by recording the ``total scattered intensity", i.e., $J_{\rm sca}=\int dt\: I_{\rm sca}(t)$, or, perfectly (ii) by simply omitting the signal before, e.g., t=25 fs in Fig.~\ref{fig3}. In this work, we study the former approach: we record the total scattered field.

 Fig~\ref{fig4} demonstrates a clear picture how the ``total near-field intensity", $J_{\rm near}({\bf r})=\int dt \: I_{\rm near}({\bf r},t)$, changes for different locations on  the tip apex. As evident from Fig.~\ref{fig4}a, total near-field intensity is 1-order of magnitude larger near and below the QO compared to other locations on tip surface. Thus, it provides a $\sim 1$ nm-size resolution. Fig.~\ref{fig4}b plots the total near-field intensity for different positions below the auxiliary QO for two different values of the oscillator strength $f=1$ and $f=0.1$~\cite{PS1}.

In Fig.~\ref{fig5}, we present the ``total scattered intensity", $J_{\rm sca}$, when the tip scans the nanoparticles, that is, for different positions of the tip-center. We calculate the $J_{\rm sca}$ in our 3D FDTD simulations as follows. We place 3D intensity monitors encapsulating only on the scanned 2-nm sized nanoparticles. Our logic is simple: the field scattered by the scanned nanoparticles is proportional to the induced near-field intensity on (and in) the nanoparticles~\cite{shalaev2006nanophotonics,PS3}. In Figs.~\ref{fig5}a and \ref{fig5}b, we scan a single 2 nm-sized nanoparticle; using a gold tip alone and and by employing the auxiliary QO, respectively. While change in the resolution between Figs.~\ref{fig5}a and \ref{fig5}b is obvious, Fig.~\ref{fig5}a presents a ``deceptive" high resolution for a 30 nm-thick gold apex. This appears because, we use a perfect (and fine-meshed) hemisphere shape for the AFM tip in the simulations while in actual experiments such a fault-free tip is almost impossible to manufacture. In an actual experiment, the stress-induced defect center, however, appears at the apex where bending is maximum~\cite{BendingDefectAPL2019}. 

Scanning two 2 nm-sized nanoparticles, separated 2 nm from each other, demonstrates the contrast between the two resolutions more explicitly. Gold-coated tip, without a QO, fails to resolve the two particles (Fig.~\ref{fig5}c) this time. Only a small wiggle between the two nanoparticles, $x=0$, appears, again, due to perfect hemisphere tip shape. A  tip employing the auxiliary QO, resolves the separation between the two nanoparticles clearly, Fig.~\ref{fig5}.

{\bf \small Complications due to oscillating tip.}--- In a SNOM experiment, the AFM tip oscillates with the natural frequency of the cantilever, typically $\Omega_{\rm tip}\sim 10^5$ Hz or less. That is, the tip does not scan a surface as if it moves along the x-direction with a constant altitude. Hence, a high repetition-rate source is required for non-sophisticated-tracking of the scattered signal with respect to the tip altitude (or tip-sample distance). This constrained the initial experiments, on ultrafast SNOM, to sources with repetition rates much larger than the $\Omega_{\rm tip}$. A recent experimental study~\cite{wang2016scattering}, however, demonstrates a method for conducting ultrafast SNOM experiments with low-repetition rate sources. Ref.~\cite{wang2016scattering} obtains the dependence of the scattered signal ($S$) on the instantaneous oscillation phase of the tip (tip-sample distance) $\Phi$, $S(\Phi)$~\cite{PS4}. The time interval we use to calculate the total scattered intensity is only 500 fs, 7-orders shorter than the oscillation frequency of the tip. Hence, the ``total scattering" data we obtain is like a point in the oscillation frame of the tip. Thus, a similar technique~\cite{wang2016scattering} can be used also in our setup in order to elect the scattering signal for an appropriate altitude of the tip.

{\bf \small In summary}, we demonstrate a novel method for obtaining sub-nm-size ultrafast SNOM imaging, which heals the current limit~($\sim$10 nm) by 1-order of magnitude. We also propose a novel technique for manufacturing an AFM (SNOM) tip in which a large oscillator strength quantum object (a defect-center) is located necessarily at the lowest-altitude of tip apex~\cite{BendingDefectAPL2019,AtatureNatureComm2017}. The method works both with low and high repetition rate sources.

Though we concentrate on on SNOM applications, our setup (Fig.~\ref{fig1}) can equally be utilized for nonlinear microscopy techniques~\cite{ye2012plasmonic,postaci2018silent,zhang2013coherent,zhang2013chemical} where small adjustments of the QO resonance is vital for Fano-enhanced phenomena to take place. Higher localization, compared to the ones using Fano resonances with dark-modes~\cite{ye2012plasmonic,zhang2013coherent}, can be achieved since a QO is much smaller than dark-mode profiles. The new method also enables ultrahigh resolution chemical manipulation on surfaces~\cite{zhan2018plasmon,kazuma2020single}.

\section*{Acknowledgment}
We gratefully thank Ceyhun Bulutay for illuminating discussions. The full-credit for the utilization of a stress-induced defect-center for the new method belongs to AB. MET acknowledges support from TUBA GEBIP 2017 and TUBITAK 1001 No:117F118. RS, AB and MET acknowledge support from TUBITAK 1001 No: 119F101.

%

\bibliography{bibliography}

\begin{thebibliography}{75}
\bibitem [{PS_()}]{PS_QO}%
  \BibitemOpen
  \href@noop {} {}\bibinfo {note} {Here, we use the word QO intentionally,
  since for such a utilization the quantized particle does not need to emit
  (fluoresce).}\BibitemShut {Stop}%
\bibitem [{PS1()}]{PS1}%
  \BibitemOpen
  \href@noop {} {}\bibinfo {note} {For the $f=0.1$ auxiliary QO, we use a 0.5
  nm thick layer of ${\rm SiO}_2$.}\BibitemShut {Stop}%
\bibitem [{PS2()}]{PS2}%
  \BibitemOpen
  \href@noop {} {}\bibinfo {note} {Here such different modes appear due to the
  perfect shape of the spherical auxiliary QO. When we increase the mesh-size,
  fringes disappear. Real-world quantum emitters are not spheres.}\BibitemShut
  {Stop}%
\bibitem [{PS3()}]{PS3}%
  \BibitemOpen
  \href@noop {} {}\bibinfo {note} {Naturally, there will be complications in
  scattering process. In this prominent work, however, we aim a basic
  demonstration of the novel phenomenon, rather than exactly simulating a
  phenomenon conducted experimentally.}\BibitemShut {Stop}%
\end{thebibliography}

\end{document}